# Three-dimensional structure of the seismo-electromagnetic ionospheric electron density disturbances

**M.I. Karpov, A.A. Namgaladze, O.V. Zolotov**
*Polytechnic Faculty of MSTU, Physics Chair*

**Abstract.** The paper presents the three-dimensional structure of the ionospheric electron density disturbances triggered by the vertical electric currents flowing between the Earth and ionosphere over the faults before the strong earthquakes. The results were obtained using the global numerical Earth's Upper Atmosphere Model (UAM). The vertical electric currents flowing between the Earth and ionosphere over the faults were used as lower boundary conditions for the UAM electric potential equation. The UAM calculated 3D structure of the ionospheric electron density disturbances demonstrates an importance of all three ionospheric plasma drift directions (movements) – vertical, meridional and zonal but not only vertical one.

**Аннотация.** В статье представлена трёхмерная структура вариаций ионосферной электронной концентрации под воздействием вертикальных электрических токов, текущих между Землёй и ионосферой над разломами перед сильными землетрясениями. Исследование проводилось с помощью глобальной трёхмерной численной нестационарной модели верхней атмосферы Земли UAM. Вертикальные токи предположительно сейсмического происхождения задавались в качестве нижних граничных условий для уравнения электрического потенциала модели. Показано, что для объяснения физического механизма формирования исследуемых аномалий следует учитывать не только вертикальный перенос плазмы под действием зональной компоненты электрического поля, но и горизонтальные (зональные и меридиональные) движения ионосферной плазмы F2-области ионосферы.

**Ключевые слова:** электронная концентрация, моделирование, UAM, землетрясение, вертикальный электрический ток, ионосфера, дрейф плазмы

## 1. Introduction

In the recent investigations (*Namgaladze*, 2010; *Namgaladze et al.*, 2011a; 2012; *Namgaladze, Zolotov*, 2011; 2012; *Karpov et al.*, 2012) we presented UAM (Upper Atmosphere Model) simulated ionosphere TEC response on the vertical seismogenic electric currents flowing between the Earth and ionosphere. Electric current parameters like magnitude, direction, latitudinal sources' position, spatial distribution were varied in these investigations. We also took into account different "back" ("return") currents, i.e. spread out over the whole globe or situated round the borders of the "straight" current. UAM simulations reproduced (1) the sunlit ionosphere income effects, i.e., TEC disturbances depression with sunrise terminator and subsolar point income and their recovery with sunset terminator arrival; (2) the magnetic conjugation of the disturbed regions; (3) the asymmetry of the seismo-related variations relative the geomagnetic equator due to different conditions (neutral composition, winds) in the Northern (winter) and Southern (summer) Hemispheres. It was also shown that (4) changes of the vertical electric currents' sign lead to "mirror-reverse" of the positive and negative TEC effects relative to the "epicenter" magnetic meridian; and (5) vertical electric currents' sources with the fixed constant density's amplitude set at 30° and 45° magnetic latitudes produced stronger TEC disturbances than at 5° and 15°; (6) spread over the whole Earth "back" ("return") currents generated TEC disturbances similar to the case with absent "back" ("return") currents; "back" ("return") currents placed round the border of the "straight" current significantly reduced the magnitude and spatially occupied area of the corresponding TEC variations. In general, the modeled TEC disturbances agreed with reported, e.g. in (*Pulinets, Boyarchuk*, 2004; *Namgaladze et al.*, 2011b; 2012; *Namgaladze, Zolotov*, 2012; *Romanovskaya et al.*, 2012), pre-earthquake TEC features' behavior. In all the named above studies we modeled seismic impact on the ionosphere by setting at the given numerical grid's nodes (composing "point"- and "line"-kind sources) constant (or fixed) additional vertical electric currents' magnitude.

In the present paper we describe the new model calculations' results obtained using the new ones – gaussoid-like – analytical expressions for the latitude-longitude density distributions of the vertical electric currents flowing between the Earth and ionosphere over the faults zones before the strong earthquakes. The obtained 3D structure of the ionospheric electron density disturbances is analyzed for the better understanding of the physical mechanism of the ionospheric electron density and total electron content disturbances observed before the strong earthquakes.

*Karpov M.I. et al. Three-dimensional structure of the seismo-electromagnetic…*

## 2. Simulation method

We have calculated the 3D spatial distribution of the ionospheric electron number density disturbances created by the vertical electric currents flowing between the Earth and ionosphere over the faults using the Upper Atmosphere Model (UAM) – global numerical model of the Earth's upper atmosphere (thermosphere, ionosphere, plasmasphere and inner magnetosphere). UAM calculates main physical parameters of the upper atmosphere such as densities, temperatures and velocities of the neutral (O, $O_2$, $N_2$, H) and charged ($O_2^+$, $NO^+$, $O^+$, $H^+$ and electrons) species of the near-Earth plasma by numerical integration of the continuity, momentum and heat balance equations as well as the equation of the electric field potential (*Namgaladze et al.*, 1988; 1991; 1998a; 1998b).

An external electric current of seismic origin flowing between the Earth and ionosphere was used as a model input for the calculations of the ionospheric electric fields and the corresponding TEC variations. These additional sources of the electric current were switched on at the lower boundary (80 km over the Earth's surface) in the UAM electric potential equation, which was solved numerically jointly with all other UAM equations for neutral and ionized gases. Quiet background conditions were obtained by performing regular calculations without any additional electric currents' sources.

*Karpov et al.* (2012) showed that "back" ("return") currents spread out all over the globe (so that total current in global circuit did not change) had no much impact on numerical results so "back" currents were not taken into consideration in the present study.

In previous calculations (*Namgaladze et al.*, 2011a; 2012; *Namgaladze, Zolotov*, 2011; 2012; *Karpov et al.*, 2012) additional electric currents with magnitude of $4 \cdot 10^{-8}$ A/m$^2$ were switched on at single or multiple grid nodes. The main difference of present calculations is that the vertical current magnitudes' spatial distributions were defined by the gaussoid-like function:

$$j(\Phi, \Lambda) = j_{max} \exp[-(\Phi - \Phi_{max})^2/(\Delta\Phi)^2 - (\Lambda - \Lambda_{max})^2/(\Delta\Lambda)^2],$$

where $\Phi$ and $\Lambda$ are the geomagnetic latitude and longitude of the vertical electric current source position, $j_{max}$ is the maximum amplitude of the electric current source switched on at the $\Phi_{max}$ geomagnetic latitude and $\Lambda_{max}$ geomagnetic longitude. $\Delta\Phi$ and $\Delta\Lambda$ define how fast the current magnitude reduces relatively to the maximum magnitude along the magnetic meridian and magnetic parallel respectively.

The main electric current source with amplitude of $j_{max} = 4 \cdot 10^{-8}$ A/m$^2$ was switched on at the grid point that corresponds to the Haiti earthquake epicenter area ($\Phi_{max} = 30°$ mag. lat., $\Lambda_{max} = 0°$ mag. long.) that happened 12th January, 2010. The grid steps over the magnetic longitude and latitude were chosen as 5° by 5°.

In order to check the dependence of seismo-electromagnetic parameters upon the spatial density distribution and total value of the electric current, the following values of $\Delta\Phi$ and $\Delta\Lambda$ were used for simulations:

1) $\Delta\Phi = 2.5°$ mag. lat., $\Delta\Lambda = 2.5°$ mag. long.;
2) $\Delta\Phi = 2.5°$ mag. lat., $\Delta\Lambda = 5.0°$ mag. long.;
3) $\Delta\Phi = 5.0°$ mag. lat., $\Delta\Lambda = 2.5°$ mag. long.;
4) $\Delta\Phi = 5.0°$ mag. lat., $\Delta\Lambda = 5.0°$ mag. long.

Corresponding for the items (1-4) patterns of the electric currents' density distribution are shown in Fig. 1. As one can see, electric currents in cases (2-3) have selected directions (along the magnetic meridian and parallel respectively).

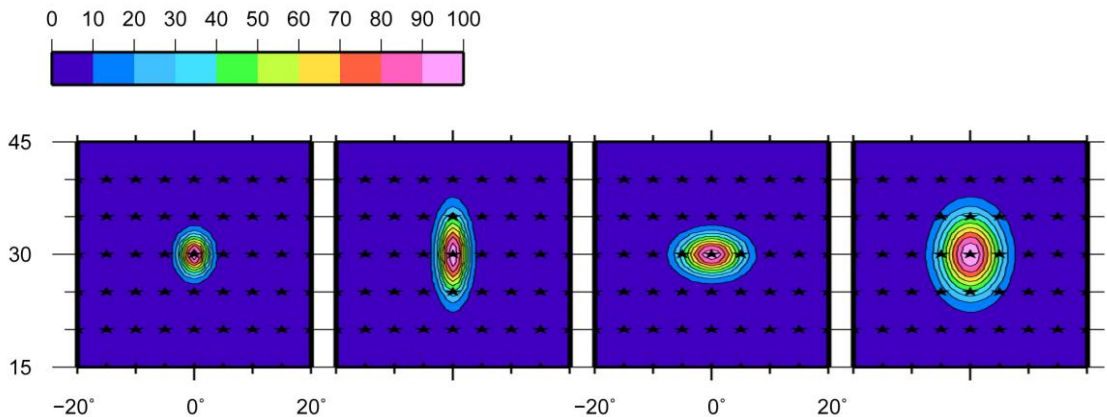

Fig. 1. The magnetic maps of the magnitudes of the electric currents $j(\Phi, \Lambda)$ relatively maximum magnitude $j_{max}$ (%) in cases 1-4 (from left to right). The stars denotes numerical grid nodes

**3. Modeled electric potential differences and TEC disturbances prior strong earthquakes**

Numerical simulation results with presented in Fig. 1 vertical electric current's density distribution are shown in Fig. 2 (longitude-latitude electric potential disturbances for 02-24UT) and Figs 3-4 (longitude-latitude TEC disturbances relative to the quiet conditions).

As one can see in Fig. 2, additional electric potential generated by external electric current increases as $\Delta\Phi$ and $\Delta\Lambda$ grow up. It reaches value of 15 kV in case (1) and 30kV in case (4) due to the total electric current increase. Electric potential disturbances reduce down to zero in all cases with approaching of the sunrise terminator.

Corresponding TEC disturbances are presented in Fig. 3 and manifest themselves as: (1) TEC depressions with well-conducting sunlit ionosphere income and restoration of the anomalies with sunset terminator coming. (2) TEC effects happen at both the EQ epicenter and magnetically conjugated areas. (3) TEC anomalously disturbed areas are asymmetric relatively to the magnetic equator due to differences in ionospheric conditions in the Northern (winter) and Southern (summer) Hemispheres. (4) Disturbances exist for all night-time period of vertical electric currents' action and (5) are linked to the epicenter area (correspond to the position of that currents). It should be noted that disturbances disappear later than electric potential difference with a time-delay of about 4-6 h due to the ionosphere inertness. (6) Electric currents of opposite direction produces stronger positive TEC disturbances and weaker negative TEC disturbances (see Fig. 4). There is a mirror reverse of negative and positive disturbances relative to the magnetic meridian crossing the epicenter and magnetic conjugated point.

The increase of additional total seismic vertical electric current in the considered region (see corresponding vertical electric current distribution in Fig. 1 and electric potentials in Fig. 2) in global circuit by increase of $\Delta\Phi$ and $\Delta\Lambda$ results in corresponding increase of the TEC disturbances. If we keep the total vertical electric current, but change the spatial configuration of the vertical electric current distribution by setting the selected spatial direction, we obtain that TEC disturbances are stronger in case of meridian-elongated source over the magnetic parallel-aligned source (cf. Fig. 3 row 2 vs row 3; also Fig. 4 row 2 vs row 3).

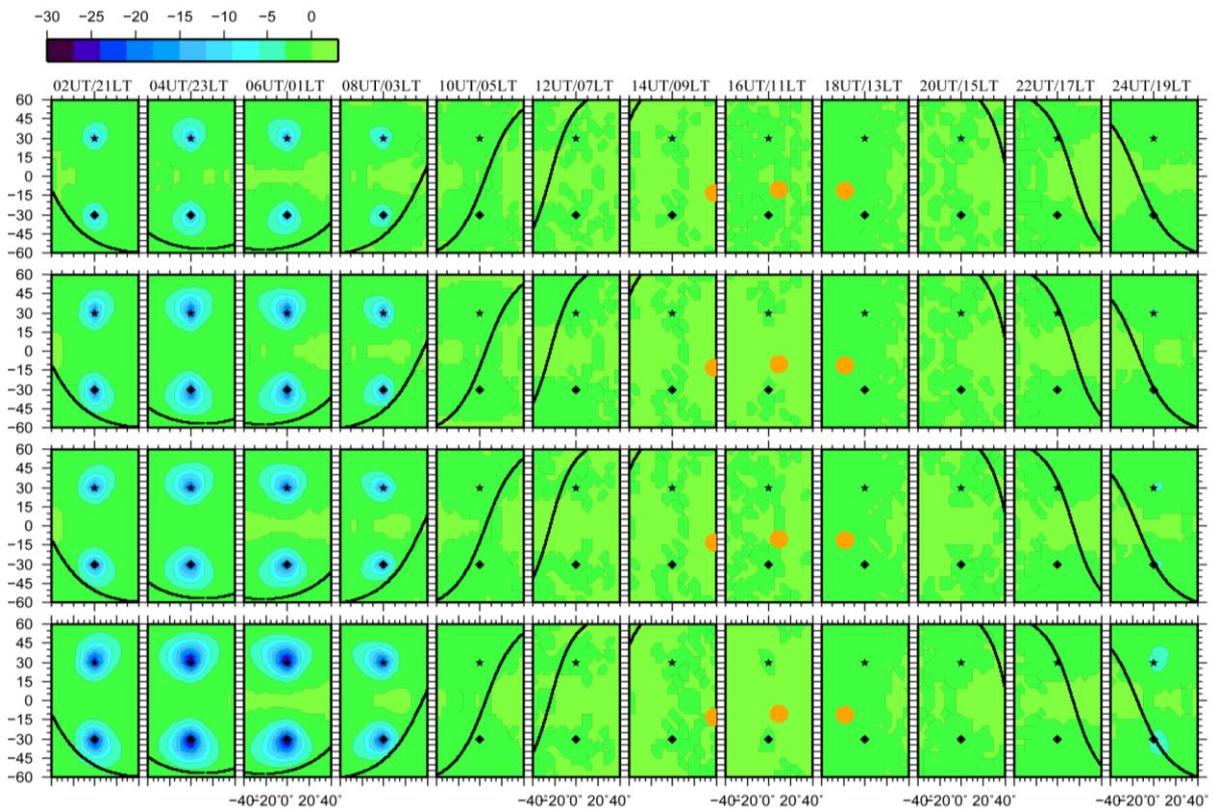

Fig. 2. The magnetic maps of calculated electric potential differences (kV) relatively background values for cases (from top to bottom): (1) $\Delta\Phi$ = 2.5° mag. lat., $\Delta\Lambda$ = 2.5° mag. long.; (2) $\Delta\Phi$ = 2.5° mag. lat., $\Delta\Lambda$ = 5.0° mag. long.; (3) $\Delta\Phi$ = 5.0° mag. lat., $\Delta\Lambda$ = 2.5° mag. long.; (4) $\Delta\Phi$ = 5.0° mag. lat., $\Delta\Lambda$ = 5.0° mag. long. Stars denotes the epicenter area; diamond denotes magnetic conjugated point. Sub-solar point and terminators positions are denoted by orange circle and black line respectively

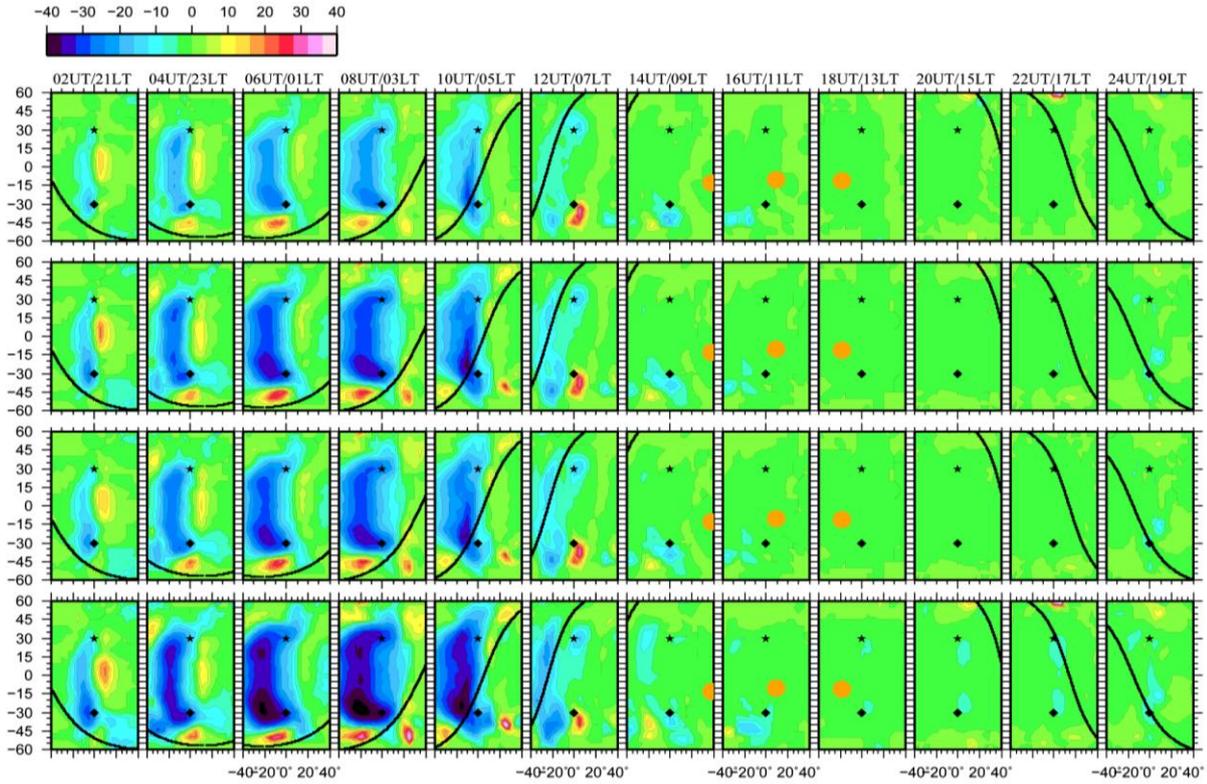

Fig. 3. The magnetic maps of calculated TEC deviations (%) relatively non-disturbed conditions for cases
(from top to bottom): (1) $\Delta\Phi = \Delta\Lambda = 2.5°$ mag. long.; (2) $\Delta\Phi = 2.5°$ mag. lat., $\Delta\Lambda = 5.0°$ mag. long.;
(3) $\Delta\Phi = 5.0°$ mag. lat., $\Delta\Lambda = 2.5°$ mag. long.; (4) $\Delta\Phi = \Delta\Lambda = 5.0°$ mag. long

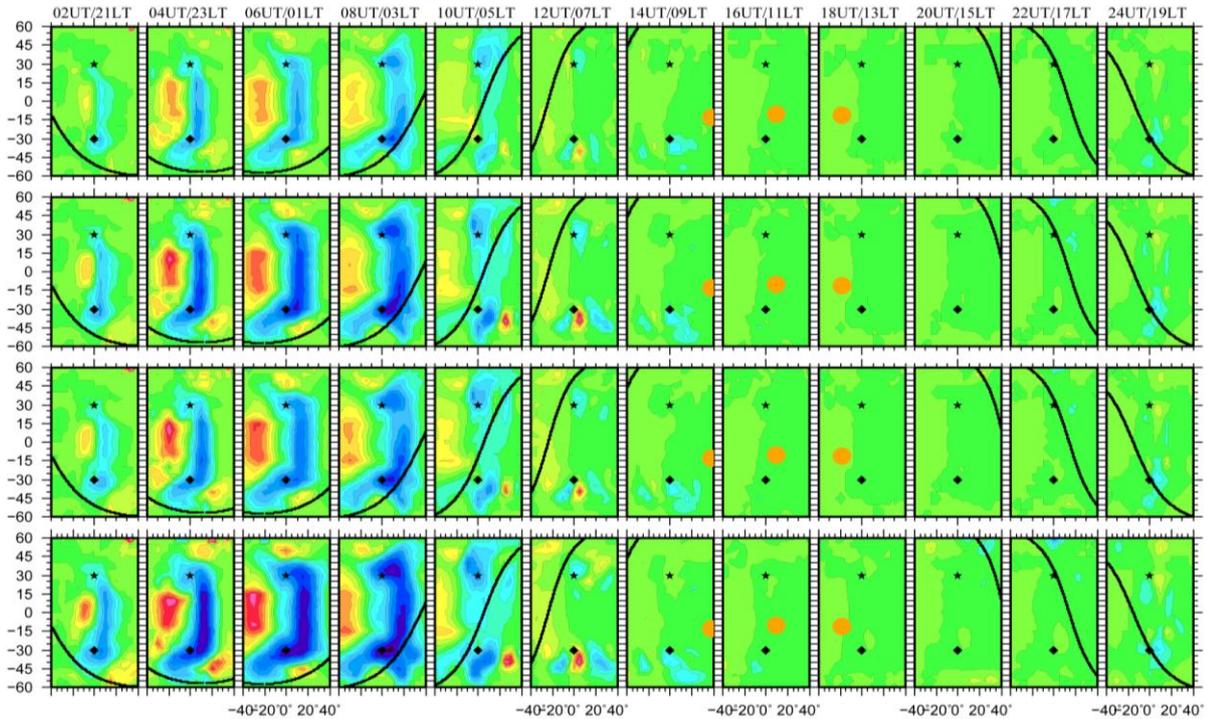

Fig. 4. Calculated TEC deviations (%) as in Fig. 2, but electric currents have the opposite
(from the Earth to the ionosphere) direction

**4. Three-dimensional structure of the ionospheric electron density disturbances**

Let us now consider three-dimensional structure of seismo-electromagnetic electron density variations along magnetic longitudes that correspond to the maximal magnitudes of TEC disturbances: +20°, 0° and -20° mag. long. The vertical electric current configuration with $\Delta\Phi = \Delta\Lambda = 5.0°$ (bottom row in Figs 1-3) are presented only.

Modeled latitude-altitude maps of the electron number density for quite conditions (without external vertical electric currents) are presented in Fig. 5. They show the Appleton equatorial anomaly evolution during the night-time at three longitudes around the EQ epicenter. We note that NmF2 and HmF2 are higher in the Southern (summer) Hemisphere than in the Northern (winter) one. This difference is a main cause of the different electron density and TEC reactions to the vertical electric current action in the Southern and Northern Hemispheres.

Numerical simulation results of the disturbed electron density are presented in Fig. 6 for the same cross-sections as in Fig. 5 but with the vertical electric current (direcred from the ionosphere to the Earth) sources (as in Fig. 1d) switched on. Their action leads to lower HmF2 values and areas of increased electron number density occupy smaller latitude-altitude region. Both crests of the Appleton anomaly shift from the geomagnetic equator. In the non-disturbed case the equatorial "trough" is deeper and stronger manifested.

Results of the similar modeling but with opposite direction of the vertical electric currents are presented in Fig. 7. We see filling of the "trough" at -20° and 0° mag. long. and deepening of the "trough" at +20° mag. long. Electron density reduces in both hemispheres till 05LT.

Calculated latitude-altitude patterns of the electron number density disturbances are presented in Fig. 8 for the vertical electric currents flowing from the ionosphere to the Earth. Electron density disturbances along 340° mag. long. (top row) reduce at altitudes of 200-300 km at the epicenter latitude (30° mag. lat.) and magnetic conjugated area and increase at the magnetic equator. Positive electron number disturbances along the epicenter longitude (0° mag. long.) shown in the middle row take place at altitudes of 200-400 km followed by negative disturbances after 23LT. Effects are stronger pronounced in the Southern Hemisphere. Disturbances along 20° mag. long. (bottom row) are much weaker than at other meridians.

Change of direction of the vertical electric currents (Fig. 9) leads to the positive electron number density disturbances along 340° mag. long. (top row) and 0° mag. long. (middle row) at the magnetic equator. Negative electron number density disturbances take place in the Northern and Southern Hemispheres along 0° and 20° mag. lat. and at the equator along 20° mag. lat. at 04-08UT.

From comparison of quite background variations with disturbed electron number density variations (Fig. 6-7), lg(Ne) differences (Fig. 8-9) and TEC disturbances (Fig. 3, bottom row) we conclude that plasma movements in all three directions (vertical, meridional and zonal) happen and are important for proper description of the electron number density 3D structure modifications under the influence of seismogenic vertical electric currents flowing between the Earth and ionosphere.

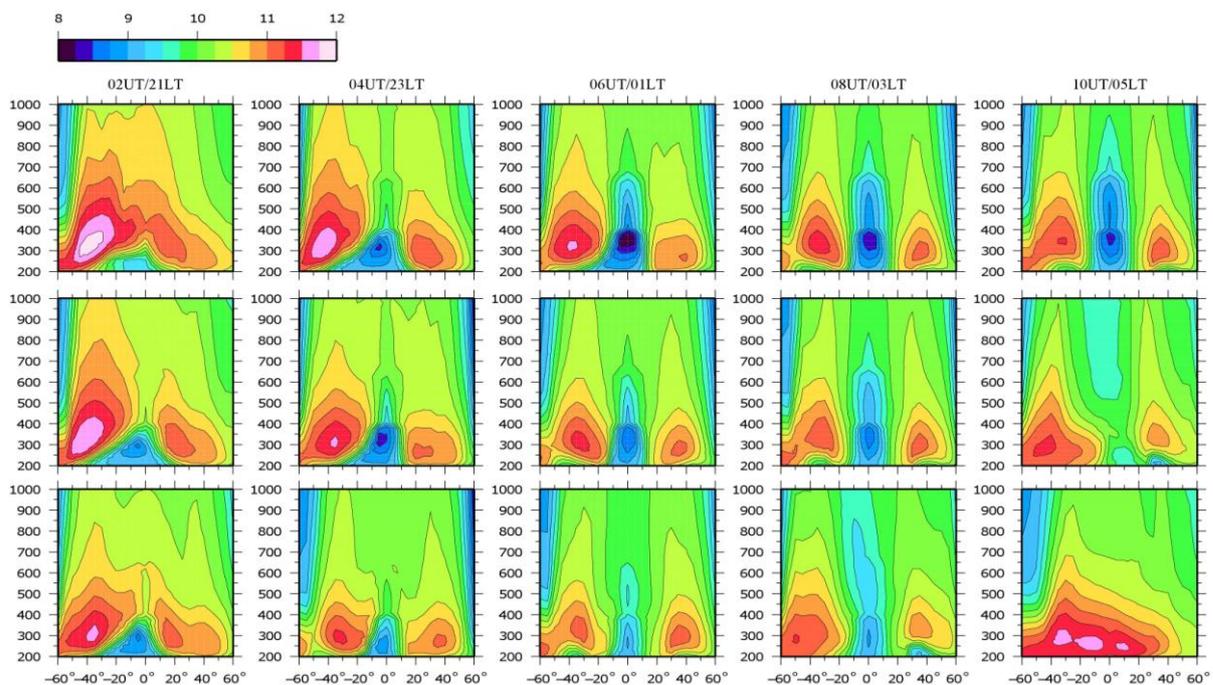

Fig. 5. Latitude-altitude maps of the electron density for quite conditions (lg(Ne)Q) at 21-05LT
along the 340° (top), 0° (middle) and 20° (bottom) mag. long.
LT labels correspond to the EQ epicenter's position

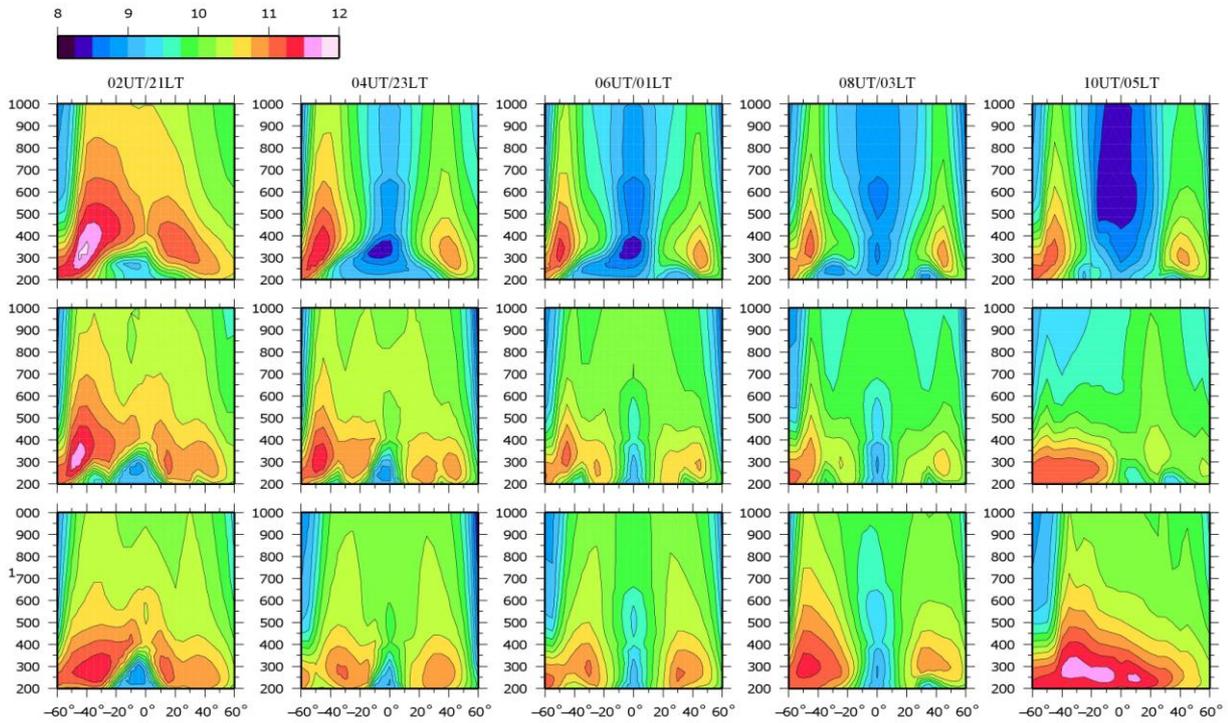

Fig. 6. Latitude-altitude maps of the disturbed electron density (lg(Ne)D) at 21-05LT
along the 340° (top), 0° (middle) and 20° (bottom) mag. long.

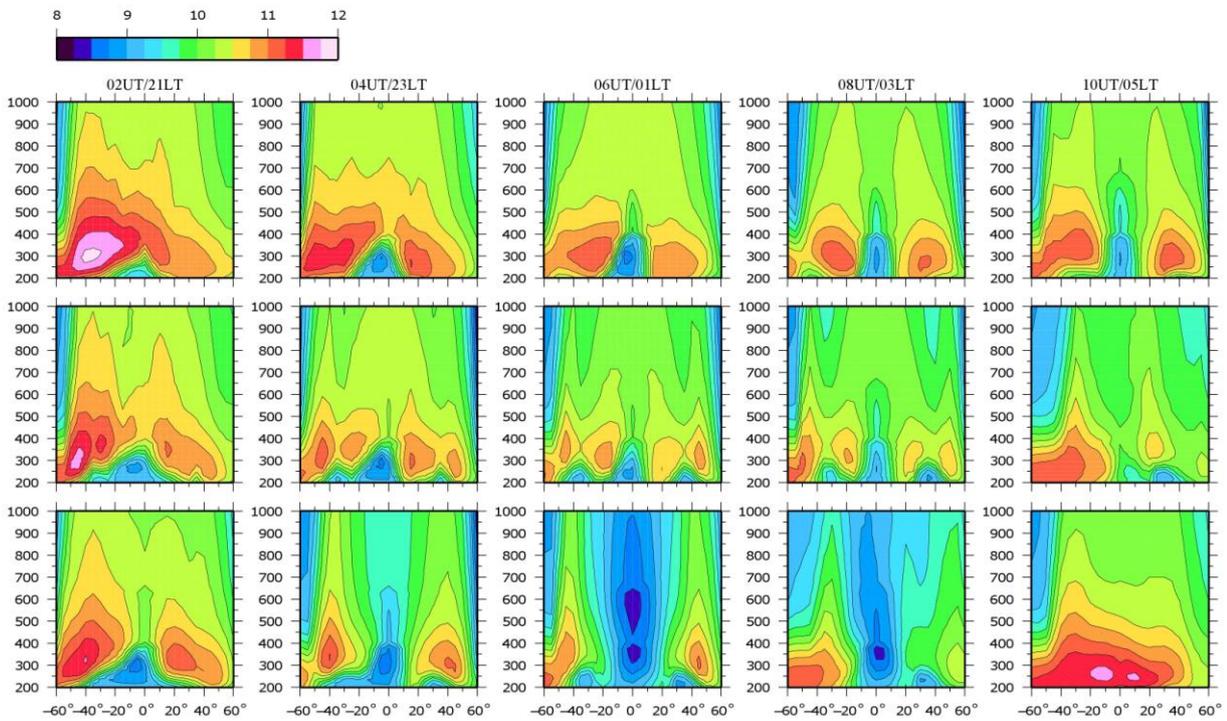

Fig. 7. The same as in Fig. 6 but with vertical electric current flowing from the Earth to the ionosphere

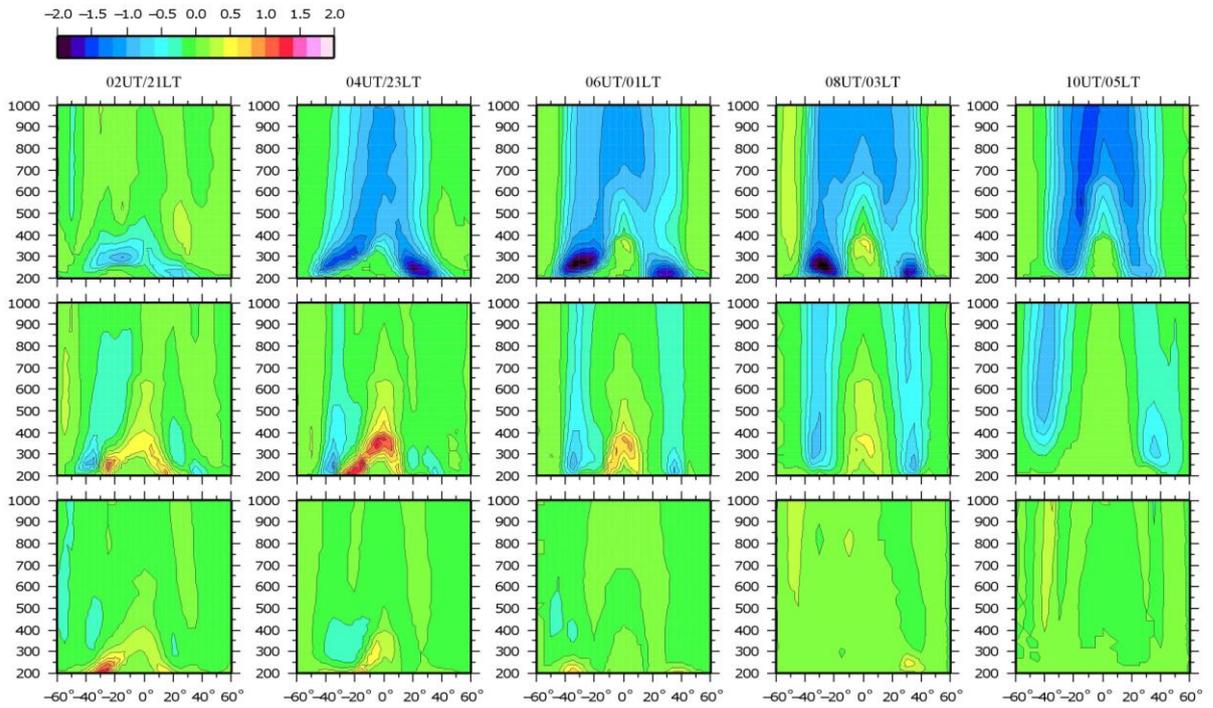

Fig. 8. Modeled latitude-altitude disturbances of the electron density relative to quiet conditions lg(Ne)D – lg(N)Q at 02-10UT along the 340° (top), 0° (middle) and 20° (bottom) mag. long. for the vertical electric currents flowing from the ionosphere to the Earth

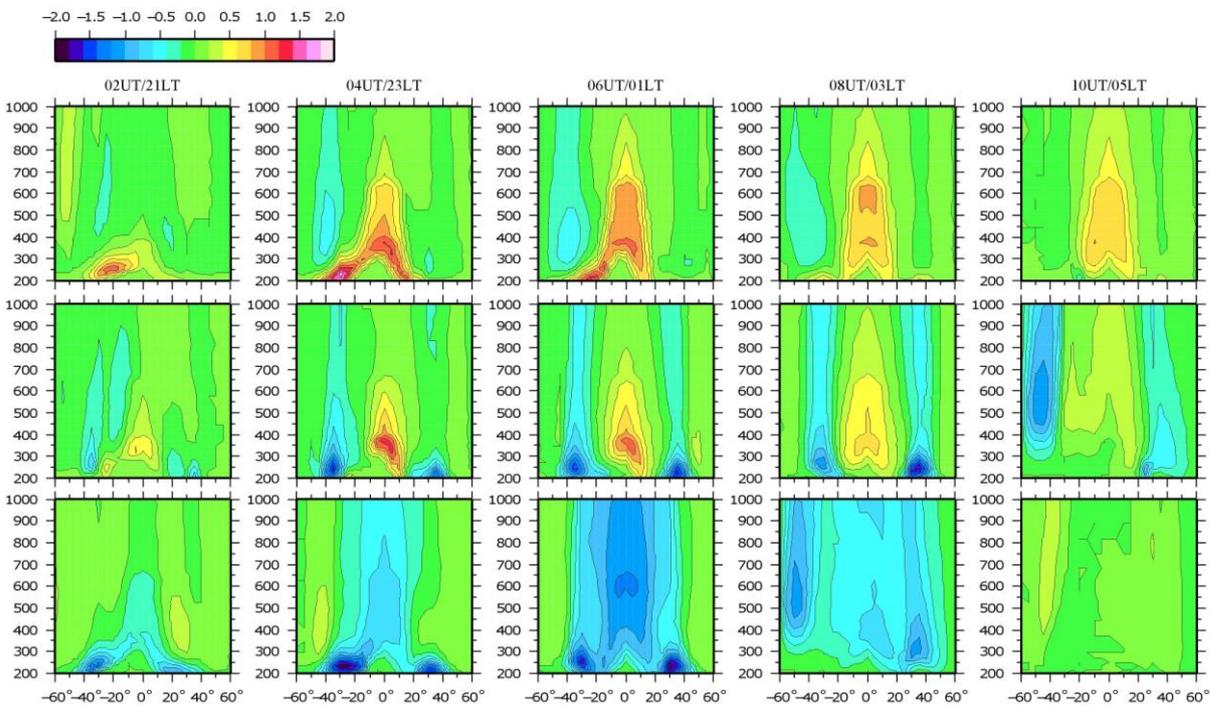

Fig. 9. The same as in Fig. 8, but the vertical electric currents have opposite direction: from the Earth to the ionosphere

## 5. Discussion

The results of the electric potential difference and corresponding TEC disturbances simulations presented in Figs 2-4 demonstrate that the presentation of the seismic vertical electric currents flowing between the Earth and ionosphere by the gaussoid-like function does not change drastically the main features revealed from the previous simulations. But its using permits more independence of the model calculation results on the mumerical grid used.

*Namgaladze et al.* (2007; 2009a; 2009b) consider that the most probable reason of the NmF2 and TEC disturbances observed before the earthquakes is the vertical drift of the F2-region ionospheric plasma under the influence of the zonal electric field of seismogenic origin. In the middle latitudes the upward electromagnetic drift, created by the eastward electric field, leads to the increase of the NmF2 and TEC due to the plasma transport to the regions with lower concentration of the neutral molecules and, consequently, with lower loss rate of dominating ions O+ in the ion-molecular reactions. The electric field of the opposite direction (westward) creates the opposite – negative – effect in NmF2 and TEC. In the low latitude regions (near the geomagnetic equator) the increase of the eastward electric field leads to the deepening of the Appleton anomaly minimum ("trough" over the magnetic equator in the latitudinal distribution of electron concentration) due to the intensification of the fountain-effect.

The ionospheric electric field created by the vertical electric currents flowing between the Earth and ionosphere is perpendicular to the geomagnetic field and has all three components (zonal, meridional and vertical) in the spherical geomagnetic coordinate system as well as the corresponding electromagnetic drift velocity which is perpendicular to both electric and geomagnetic fields. We should keep in mind that the ionospheric plasma beside of the drift motion participates simultaneously in the diffusion along the geomagnetic field restoring the balance between the plasma pressure gradient and gravitational forces distorted by the drift. All these motions (drift and diffusion as well as neutral wind action) influence on the ionospheric plasma density redistribution when the vertical electric currents are switched on. All these processes are taken into account in the UAM and form the electron density and TEC disturbances presented above.

3D structures of the electron number density presented in Figs 5-6 demonstrate that although vertical F2-layer ionospheric plasma drift under the influence of the zonal component of the disturbed electric field plays major role in the TEC disturbance formation, horizontal movements (meridional and zonal components of plasma drift) should not be neglected.

## 6. Conclusions

Three-dimensional structure of the ionospheric electron density disturbances created by the vertical electric currents flowing between the Earth and ionosphere over the faults before the strong earthquakes have been calculated using the global numerical Upper Atmosphere Model. The vertical electric currents flowing between the Earth and ionosphere over the faults were used as input lower boundary conditions for the UAM electric potential equation. The vertical current magnitudes' spatial distributions were defined by the gaussoid-like function.

Main phenomenological features of TEC disturbances revealed in previous simulations were reproduced and their dependences on the vertical electric current parameters were established.

The UAM calculated 3D structure of the ionospheric electron density disturbances demonstrates importance of all three F2-region ionospheric plasma drift components in producing NmF2 and TEC disturbances prior strong earthquakes. Not only vertical but also meridional and zonal plasma drift movements should be taken into account.

## References

**Karpov M.I., Zolotov O.V., Namgaladze A.N.** Modeling of the ionosphere response on the earthquake preparation. *Proc. of the MSTU*, v.15, N 2, p.471-476, arXiv: 1205.0415, 2012.

**Namgaladze A.A., Korenkov Yu.N., Klimenko V.V., Karpov I.V., Bessarab F.S., Surotkin V.A., Glushchenko T.A., Naumova N.M.** Global model of the thermosphere-ionosphere-protonosphere system. *Pure and Applied Geophysics*, v.127, N 2/3, p.219-254, doi:10.1007/BF00879812, 1988.

**Namgaladze A.A., Korenkov Yu.N., Klimenko V.V., Karpov I.V., Surotkin V.A., Naumova N.M.** Numerical modeling of the thermosphere-ionosphere-protonosphere system. *Journal of Atmospheric and Terrestrial Physics*, v.53, N 11/12, p.1113-1124, doi:10.1016/0021-9169(91)90060-K, 1991.

**Namgaladze A.A., Martynenko O.V., Volkov M.A., Namgaladze A.N., Yurik R.Yu.** High-latitude version of the global numeric model of the Earth's upper atmosphere. *Proc. of the MSTU*, v.1, N 2, p.23-84, URL: http://goo.gl/8x9f2, 1998a.

**Namgaladze A.A., Martynenko O.V., Namgaladze A.N.** Global model of the upper atmosphere with variable latitudinal integration step. *Int. J. of Geomagnetism and Aeronomy*, v.1, N 1, p.53-58, 1998b.

**Namgaladze A.A., Shagimuratov I.I., Zakharenkova I.E., Zolotov O.V., Martynenko O.V.** Possible mechanism of the TEC enhancements observed before earthquakes. *XXIV IUGG General Assembly, Perugia, Italy, 02-13 July 2007, Session JSS010*, 2007.

**Namgaladze A.A., Klimenko M.V., Klimenko V.V., Zakharenkova I.E.** Physical mechanism and mathematical modeling of earthquake ionospheric precursors registered in total electron content. *Geomagnetism and Aeronomy*, v.49, N 2, p. 252-262, doi:10.1134/S0016793209020169, 2009a.

**Namgaladze A.A., Zolotov O.V., Zakharenkova I.E., Shagimuratov I.I., Martynenko O.V.** Ionospheric total electron content variations observed before earthquakes: Possible physical mechanism and modeling. *Proc. of the MSTU*, v.12, N 2, p.308-315, ArXivID: 0905.3313 URL: http://goo.gl/A8cLx, 2009b.

**Namgaladze A.A.** Physical model of earthquake ionospheric precursors (Invited). *Abstract NH24A-03 presented at 2010 Fall Meeting,* AGU, San Francisco, Calif., 13-17 Dec., 2010.

**Namgaladze A.A., Zolotov O.V.** Ionospheric effects from different seismogenic electric field sources. *XXXth URSI General Assembly and Scientific Symposium,* Istanbul, Turkey. IEEE, August. doi:10.1109/URSIGASS.2011.6051040, 2011.

**Namgaladze A.A., Zolotov O.V., Prokhorov B.E.** Ionospheric TEC effects related to the electric field generated by external electric current flowing between faults and ionosphere. *Physics of Auroral Phenomena: Abstracts of 34 Annual Seminar, Apatity, March 1-4, 2011, Apatity*, p.50, 2011a.

**Namgaladze A.A., Zolotov O.V., Prokhorov B.E.** The TEC signatures as strong seismic event precursors. *XXXth URSI General Assembly and Scientific Symposium. Istanbul, Turkey. IEEE, August*, doi:10.1109/URSIGASS.2011.6051048, 2011b.

**Namgaladze A.A., Ferster M., Prokhorov B.E., Zolotov O.V.** Electromagnetic drivers in the upper atmosphere: Observations and modeling. *In: "The Atmosphere and Ionosphere: Elementary Processes, Discharges and Plasmoids (Physics of Earth and Space Environments)", Eds. Bychkov V., Golubkov G., Nikitin A., Springer*, 300 p., 2012.

**Namgaladze A.A., Zolotov O.V.** Ionospheric effects of seismogenic disturbances of the global electric circuit seismogenic disturbances. *In: "Earthquakes: Triggers, Environmental Impact and Potential Hazards",* Kostas Konstantinou (Ed.), NovaPub, 2012.

**Pulinets S.A., Boyarchuk K.** Ionospheric precursors of earthquakes. *Springer, Berlin, Germany*, 315 p., 2004.

**Romanovskaya Y.V., Namgaladze A.A., Zolotov O.V., Starikova N.A., Lopatiy V.Z.** Searching for seismo-ionospheric earthquakes precursors: Total Electron Content disturbances before 2005-2006 seismic events. *Proc. of the MSTU*, v.15, N 2, p.477-481, arXiv: 1205.0419, 2012.